\documentclass[aps,prl,twocolumn,groupedaddress,showpacs]{revtex4}

\usepackage{graphicx}
\usepackage{amssymb}

\begin{document}

\title{Shannon Entropy as Characterization Tool in Acoustics}
\author{Helios Sanchis-Alepuz}
\affiliation{Wave Phenomena Group, Department of Electronic Engineering, Polytechnic University of Valencia, C/ Camino de Vera s/n, E-46022 Valencia, Spain.}
\author{Jos\'e S\'anchez-Dehesa}
\email[E-mail:]{jsdehesa@upvnet.upv.es}
\affiliation{Wave Phenomena Group, Department of Electronic Engineering, Polytechnic University of Valencia, C/ Camino de Vera s/n, E-46022 Valencia, Spain.}
\date{\today}

\begin{abstract}
We introduce Shannon's information entropy to characterize the avoided crossing appearing in the resonant Zener-like phenomenon in ultrasonic superlattices made of two different fluidlike metamaterials. We show that Shannon's entropy gives a correct physical insight of the localization effects taking place and manifest the informational exchange of the involved acoustic states in the narrow region of parameters where the avoided crossing occurs. Results for ultrasonic structures consisting of alternating layers of methyl-metacrylate and water cavities, in which the acoustic Zener effect were recently demonstrated, are also reported. 
\end{abstract}

\maketitle

Shannon's informational entropy\cite{shannon48} boosted an increasing number of applications in physics in the last two decades.  Among others, the studies on novel and counter-intuitive ways of processing and transmitting (quantum) information are perhaps the more fascinating\cite{nielsen,proceedings}.
  The recent applications to crystallography\cite{diamond63,menendez06} and atomic physics\cite{gonzalezPRL03,gonzalezCPL03} are also examples where Shannon entropy opened new ways of interpretation of well known physical phenomena.

The aim of this Letter is to introduce Shannon's informational entropy in acoustics and to use it as a new characterization tool in problems where localization plays a fundamental role. 
Particularly, it will be employed to study the dynamics of the recently discovered\cite{sanchisPRL07} acoustical analogue of the electronic Zener effect\cite{zener}, a phenomenon that was previously observed in semiconductor superlattices\cite{schneider,rosam}. This is done by analyzing the avoided crossing occurring between acoustic levels belonging to different minibands in ultrasonic superlattices. 
Here, we introduce the formula for the Shannon entropy in acoustics and report an in-depth analysis for the case of a multilayer of two fluidlike metamaterials where both levels, their frequencies and their Shannon entropies will be discussed. Results for a multilayer of water cavities and methyl-metacrylate (plexiglas) layers, that is the structure studied in [\onlinecite{sanchisPRL07}], will be also presented to show its results for an actual system. 

Shannon entropy has been defined in atomic physics by $S_{\rho}=-\int{\rho\left({\bf r}\right){\rm ln}\rho\left({\bf r}\right)d{\bf r}}$, where $\rho\left({\bf r}\right)=\left|\psi({\bf r})\right|^2$ is the probability density distribution of a given electronic state. In acoustics, however, there is not an equivalent magnitude having an interpretation of a probability distribution. To overcome such drawback we have exploited the analogy between electronic states in quantum mechanics and acoustic levels in acoustic structures. For example, slabs of a solid material in air sustain vibrations that, like their quantum counterparts (the atomic levels), are quasibound. 
Thus, sound resonances in slabs made of cylindrical rods have been recently demonstrated\cite{sanchisPRB03}, the associated displacement fields, $u(r)$, are characterized by their localization in some spacial region and they decay after some delay time if they are excited. 
Following the analogy with atomic systems, we introduce the following probability distribution function:
\begin{equation}
{\cal{P}}({\bf r})=\left|u({\bf r})\right|^2/\int{\left|u({\bf r})\right|^2d{\bf r}},
\end{equation}
which is obtained by normalizing the square of the displacement field of a given acoustic level.
It will play in acoustics the same role that the electronic density distribution in quantum mechanics.
Therefore, the Shannon's information entropy is defined by:
\begin{equation}
S_u=-\int{{\cal{P}}({\bf r})}{\rm ln}{\cal{P}}({\bf r}){\rm d}{\bf r}
\end{equation}   
This quantity is an information measure of the spatial delocalization of the sound level in the corresponding acoustic system and, therefore, gives the uncertainty of the localization of sound. 
This quantity, like the the Shannon formula\cite{shannon48}, increases with increasing uncertainty (i.e., spreading of the displacement field of the acoustic state). 
In what follows we use $S_u$ to get a physical insight of the dynamics of an acoustic system in which two interacting acoustic levels present an avoided crossing region. The repulsion of the acoustic modes (as the external field adiabatically changes) illustrates how the avoided crossing phenomenon is a mechanism for sound localization reordering with frequency. 

Let us assume an acoustic superlattice made by stacking two-types of fluidlike materials A and W. 
In particular, we consider a multilayer made of $m$ coupled water cavities $W_m$ enclosed by $m+1$ slabs A made of an acoustic metamaterial consisting of a two-dimensional periodic distribution of rigid rods that are also embedded in water. The feasibility of this kind of metamaterials has been recently demonstrated\cite{torrentPRL06} and their acoustic parameters (density and sound velocity) can be tailored with practically no limitation\cite{torrentPRB06,torrentNJP07}.
\begin{figure}[ht]
\includegraphics[width=0.40\textwidth]{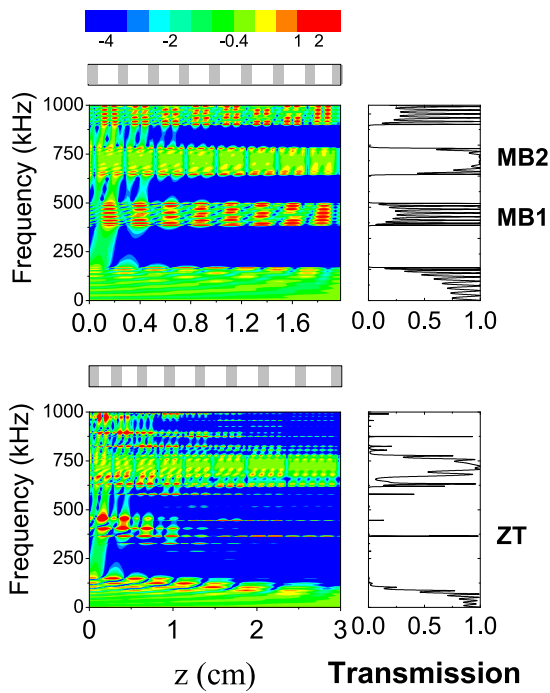}
\caption{(color online) Transfer-matrix calculation for the square of the displacement field amplitude $\left|u(z)\right|^2$ of the sound inside the acoustic superlattice schematically shown on top of each plot. Left panels represent the $log\left|u(z)\right|^2$ versus the frequency in the color scale shown at the top. Right panels plot the transmission spectra trough the entire structures.}
\label{fig:TM}
\end{figure}
Transfer matrix calculations\cite{bre} of the transmission coefficient through the multilayer as a function of the frequency and length of the multilayer have been performed by using the following inputs.
As parameter of the effective fluid associated to the A slabs we have taken $\rho_A=$5$\rho_W$, and $c_A=$0.78$c_W$, where $\rho_W(=$1 g/cm$^{3}$) and $c_W(=$1.48$\times$10$^5$cm/s) are the density and sound velocity in water, respectively.
The thicknesses of A layers are taken all equal to 0.08 cm, while those of water cavities are, for the case of the perfect superlattice equal to $d_W$=2$d_A$. 
Upper panels in Fig. \ref{fig:TM} report the calculate amplitude of the displacement field, $log\left|u(z)\right|^2$, inside a superlattice made of eight water cavities (left panel) and the corresponding transmission spectrum trough the complete structure. Two minibands, MB1 and MB2, are clearly formed. Miniband MB1 is approximately centered at the frequency corresponding to the first Fabry-Perot resonance of a water cavity with thickness $d_w$ (i.e., at $\pi c_w/d_w$=462 kHz). Thus, note how the modes of MB1 are strongly localized in the water cavities. However, note that modes in MB2 are mixed and their spatial localization is not clearly defined.
\begin{figure}[ht]
\includegraphics[width=0.40\textwidth]{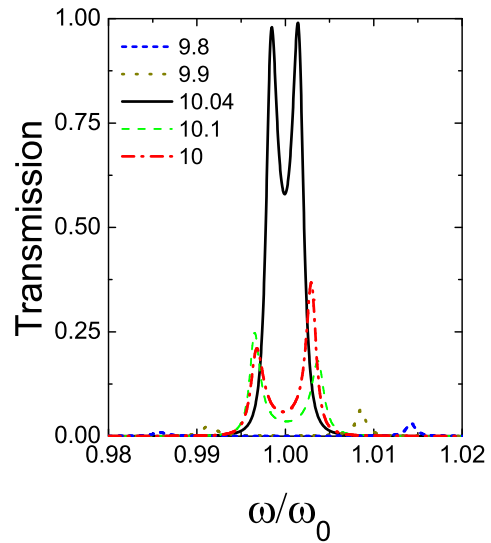}
\caption{(color online) Transfer matrix calculation of the transmission spectra of the acoustic multilayer studied in Fig. 1 around the gradient where the Zener-like resonant effect occurs. The transmission is plotted for several gradients (in $\%$) as a function of the reduced frequency, where $\omega_0$ is the central frequency at each gradient.}
\label{fig:figure2}
\end{figure} 
To observe the resonant Zener-"tunneling" (ZT) effect in an acoustic superlattice, we follow the procedure already discussed in [\onlinecite{sanchisPRL07}] and broke the translational symmetry by introducing a gradient in the thicknesses of water cavities. The gradient plays the role of a driven force producing effects similar than the electric field does in an electronic superlattice. The magnitude of the gradient is given by the dimensionless parameter 
\begin{equation}
\Delta (1/d_W)=[(1/d_{W_\ell})-(1/d_{W_{\ell-1}})]/(1/d_{W_1}),
\end{equation}
where the thickness of the first cavity ($\ell =$1) being $d_{W_1}=$0.08 cm.

Lower panels in Fig. \ref{fig:TM} show the amplitude, $log\left|u(z)\right|^2$, and the total transmission through a structure with the critical gradient $\Delta (1/d_W)_c=$10.04$\%$; i.e., that at which the acoustic ZT-like effect takes place. 
For this gradient the interaction between the higher level in MB1 and the lower level in MB2 is the strongest and the total transmission trough the structure is maximum. 
An in deep analysis of this phenomenon is shown in Fig. \ref{fig:figure2} where the transmission is plotted for several gradients around the critical. Note, how the peaks in the transmission spectra are quite small for gradients smaller or larger than the critical. Also note that the transmission profile at the critical gradient shows a double peak, which indicates that two anticrossing levels are involved in the observed phenomenon. 

\begin{figure}[h]
\includegraphics[width=0.40\textwidth]{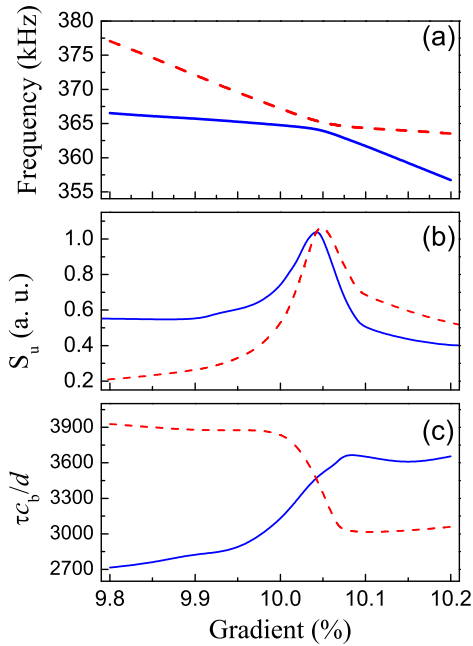}
\caption{(Color online) (a) The frequency of the interacting modes near the avoided crossing region for the system described in Fig. 1. (b) The Shannon entropy, $S_u$. (c) The lifetime, $\tau$, (in reduced units) of modes involved in the interaction process. The blue continuous (red dashed) lines represent magnitudes associated to the lower (upper) level.}
\label{fig:shannonMT}
\end{figure}

The frequency and Shannon entropy of the levels involved in the avoided crossing are represented in Fig. \ref{fig:shannonMT}(a) and \ref{fig:shannonMT}(b), respectively. For small gradients, the computed result of entropy corresponding to the lower level, $S_{u1}$, indicate that the sound in this mode is more localized than in the upper level, where its corresponding entropy $S_{u2}$ is lower. For increasing gradients both modes have a similar behavior up to the region between 9.95$\%$ and 10.06$\%$, where the modes strongly mix up and present a maximum at the gradient of 10.04$\%$.
Also note that the entropy difference $\Delta S=S_{u2}-S_{u1}$ is negative for low gradients. However, as the gradient approaches the critical value  the entropies of both levels strongly increase but their difference decreases and get the value $\Delta S=$0.022 a.u. at the critical gradient $\Delta(1/d_W)_c$. Then, for $\Delta(1/d_W) > \Delta(1/d_W)_c$ the entropies monotonically decreases and their difference enhances and its sign change to negative. This behavior is a clear and quantitative indication showing that the information-theoretic character of both levels has been exchanged when passing through the avoided crossing region. In other words, the localization properties of the lower and upper states has been exchanged when the driven force $\Delta(1/d_W)$ has been adiabatically changed between 9.95$\%$ and 10.06$\%$.
To further support this interpretation, we have calculated the lifetime of the resonances (acoustic modes) involved in the resonant phenomenon. By following the procedure described in [\onlinecite{sanchisPRB03}], we have calculated the complex frequency of the modes: $\nu_i={\rm Re}(\nu_i)+i{\rm Im}(\nu_i)$, $i=$1,2. The resonance lifetime, $\tau$ is related with the imaginary part, and the calculated values are represented in Fig. \ref{fig:shannonMT}(c) in units of $d/c_b$. 
The results support the interpretation obtained from the Shannon entropies. Longer lifetime (i.e., stronger spatial localization) is obtained for the upper mode (red dashed lines) at low gradients. Also, the lifetimes of both levels suddenly change around the critical gradient and then, after passing through the avoided crossing region, their relative magnitude has been exchanged; the lower level has longer lifetime that the upper. On the other hand, the resonance frequencies (the real part) exactly agree with the frequencies at which the peaks in the transmission spectra are observed [see Fig. \ref{fig:shannonMT}(a)]. 

\begin{figure}[ht]
\includegraphics[width=0.40\textwidth]{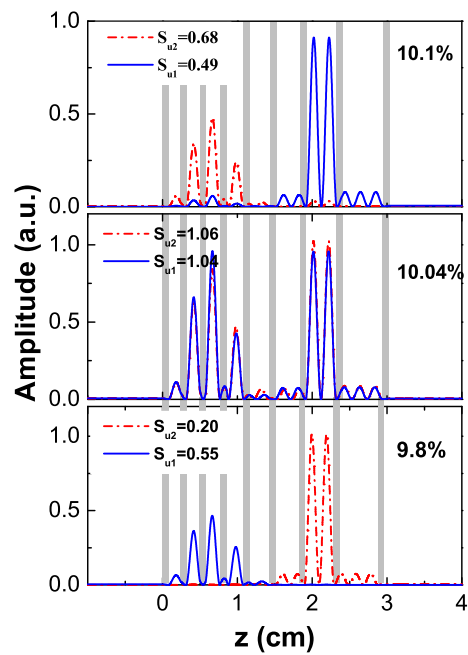}
\caption{(color online) Amplitude of acoustic modes involved in the avoided crossing effect shown in Fig. 2. The modes are plotted for several gradients around that where the avoided crossing occurs (see Fig. 1). The shadowed regions represent the metamaterial layers in the superlattice.}
\label{fig:fig4}
\end{figure}

A physical insight of the localization phenomena taking place across the critical gradient is shown in Fig.\ref{fig:fig4} where the amplitudes of both acoustic modes are plotted for three different gradients, one being the critical gradient. It is clearly demonstrated that greater values of Shannon entropy means more spreading of the acoustic modes. Moreover, at the critical gradient (10.04$\%$) the modes resulting from the interaction are the bonding and antibonding combinations of the non-interacting modes, which explain the similar spreading shown by both "wavefunctions" in Fig.\ref{fig:fig4}. 

For the comprehensiveness of our analysis we have also studied the case of the Zener-like effect experimentally characterized in a superlattice made of plexiglas and water cavities\cite{sanchisPRL07}.
 The computations have been performed with the same set of parameters reported in [\onlinecite{sanchisPRL07}]. 
 The results for the frequency, the Shannon entropy and the lifetime of the pair of levels are shown in Fig. 5. From Fig. 5(b) we observe once more that the entropy of each level get a maximum value at the gradient where the minimum distance between level is achieved. Also, the entropies maximize when passing across the critical gradient, and the relative character of both modes is exchanged.  This is corroborated by the modes lifetime shown in Fig. 5(c). Let us point out again that the main feature of Shannon entropy is that maximizes its value at the gradient 9.93$\%$, which is the critical gradient at which maximum interaction between levels is achieved. At that critical value the corresponding lifetimes of the non-crossing levels (bonding and antibonding) crosses. 

\begin{figure}[ht]
\includegraphics[width=0.40\textwidth]{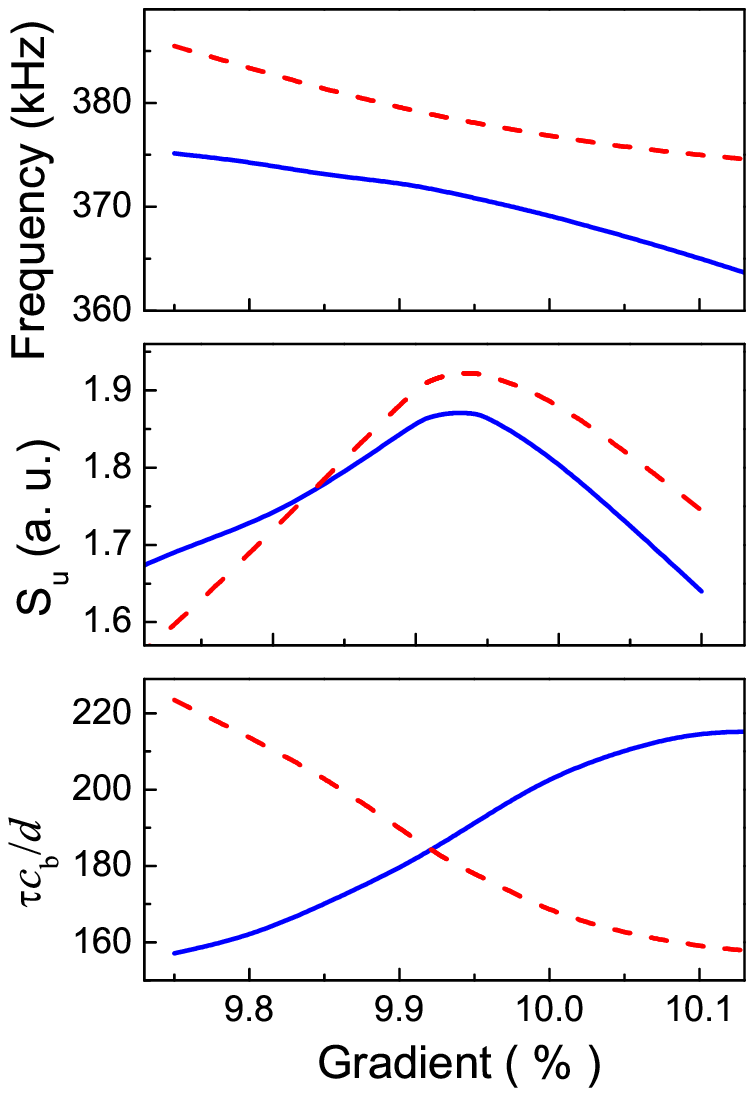}
\caption{(Color online) (a) The frequency of the interacting modes near the avoided crossing region for the system experimentally studied in [\onlinecite{sanchisPRL07}]. (b) The Shannon entropy, $S_u$, in arbitrary units. (c) The lifetime, $\tau$, (in reduced units) of modes involved in the interacting process. The blue continuous (red dashed) lines represent magnitudes associated to the lower (upper) levels.}
\label{fig:figura5}
\end{figure}

In summary, we have introduced the Shannon informational entropy in acoustics and have shown that it can be employed as an useful indicator of localization effects taking place in acoustical phenomena, like the resonant Zener-like effect. The exchange of the informational character observed between levels in this acoustical phenomenon is a translation into acoustics of the character exchange previously pointed out by von Neumann and Wigner\cite{vonnewmann29} studying interacting levels in quantum mechanics. Then, the Shannon entropy seems to be an appropriated magnitude to quantitatively estimate such character exchange. Let us stress that Shannon entropy can be experimentally determined in any acoustical system where direct measurement of displacement fields can be performed, like, for example, the structures described in [\onlinecite{Gutierrez}]. Finally, we expect that this work stimulates further applications of the Shannon entropy formula here employed, which might help in alternative interpretations in acoustics. Moreover, the application of Shannon entropy in studying localization phenomena in optics and, particularly, localization effects produced by photonic crystals is in order, since localization in these structures still remain an open problem.

\begin{acknowledgments}
Work partially supported by the Spanish MEC (No. TEC2004-03545) and by Generalitat Valenciana. H. S.-A. acknowledges a grant paid by MEC. We also thank R. Gonzal\'ez-F\'erez, Jesus S. Dehesa, and Yu. A. Kosevich for useful discussions.
\end{acknowledgments}

\end{document}